\documentstyle[12pt]{article}

\def\hybrid{\topmargin -20pt    \oddsidemargin 0pt
	\headheight 0pt \headsep 0pt
	\textwidth 6.25in       
	\textheight 9.5in       
	\marginparwidth .875in
	\parskip 5pt plus 1pt   \jot = 1.5ex}

\hybrid

\def\baselinestretch{1.2}

\catcode`\@=11

\def\marginnote#1{}
\newcount\hour
\newcount\minute
\newtoks\amorpm
\hour=\time\divide\hour by60
\minute=\time{\multiply\hour by60 \global\advance\minute by-\hour}
\edef\standardtime{{\ifnum\hour<12 \global\amorpm={am}%
	\else\global\amorpm={pm}\advance\hour by-12 \fi
	\ifnum\hour=0 \hour=12 \fi
	\number\hour:\ifnum\minute<10 0\fi\number\minute\the\amorpm}}
\edef\militarytime{\number\hour:\ifnum\minute<10 0\fi\number\minute}

\def\draftlabel#1{{\@bsphack\if@filesw {\let\thepage\relax
   \xdef\@gtempa{\write\@auxout{\string
      \newlabel{#1}{{\@currentlabel}{\thepage}}}}}\@gtempa
   \if@nobreak \ifvmode\nobreak\fi\fi\fi\@esphack}
	\gdef\@eqnlabel{#1}}
\def\@eqnlabel{}
\def\@vacuum{}
\def\draftmarginnote#1{\marginpar{\raggedright\scriptsize\tt#1}}

\def\draft{\oddsidemargin -.5truein
	\def\@oddfoot{\sl preliminary draft \hfil
	\rm\thepage\hfil\sl\today\quad\militarytime}
	\let\@evenfoot\@oddfoot \overfullrule 3pt
	\let\label=\draftlabel
	\let\marginnote=\draftmarginnote
   \def\@eqnnum{(\theequation)\rlap{\kern\marginparsep\tt\@eqnlabel}%
\global\let\@eqnlabel\@vacuum}  }

\def\preprint{\twocolumn\sloppy\flushbottom\parindent 2em
	\leftmargini 2em\leftmarginv .5em\leftmarginvi .5em
	\oddsidemargin -.5in    \evensidemargin -.5in
	\columnsep .4in \footheight 0pt
	\textwidth 10.in        \topmargin  -.4in
	\headheight 12pt \topskip .4in
	\textheight 6.9in \footskip 0pt
	\def\@oddhead{\thepage\hfil\addtocounter{page}{1}\thepage}
	\let\@evenhead\@oddhead \def\@oddfoot{} \def\@evenfoot{} }

\def\numberbysection{\@addtoreset{equation}{section}
	\def\theequation{\thesection.\arabic{equation}}}

\def\underline#1{\relax\ifmmode\@@underline#1\else
	$\@@underline{\hbox{#1}}$\relax\fi}

\def\titlepage{\@restonecolfalse\if@twocolumn\@restonecoltrue\onecolumn
     \else \newpage \fi \thispagestyle{empty}\c@page\z@
	\def\thefootnote{\fnsymbol{footnote}} }

\def\endtitlepage{\if@restonecol\twocolumn \else \newpage \fi
	\def\thefootnote{\arabic{footnote}}
	\setcounter{footnote}{0}}  

\catcode`@=12
\relax

\def\figcap{\section*{Figure Captions\markboth
	{FIGURECAPTIONS}{FIGURECAPTIONS}}\list
	{Figure \arabic{enumi}:\hfill}{\settowidth\labelwidth{Figure 999:}
	\leftmargin\labelwidth
	\advance\leftmargin\labelsep\usecounter{enumi}}}
 \relax
\def\tablecap{\section*{Table Captions\markboth
	{TABLECAPTIONS}{TABLECAPTIONS}}\list
	{Table \arabic{enumi}:\hfill}{\settowidth\labelwidth{Table 999:}
	\leftmargin\labelwidth
	\advance\leftmargin\labelsep\usecounter{enumi}}}
 \relax
\def\reflist{\section*{References\markboth
	{REFLIST}{REFLIST}}\list
	{[\arabic{enumi}]\hfill}{\settowidth\labelwidth{[999]}
	\leftmargin\labelwidth
	\advance\leftmargin\labelsep\usecounter{enumi}}}
 \relax
\makeatletter
\newcounter{pubctr}
\def\publist{\@ifnextchar[{\@publist}{\@@publist}}
\def\@publist[#1]{\list
	{[\arabic{pubctr}]\hfill}{\settowidth\labelwidth{[999]}
	\leftmargin\labelwidth
	\advance\leftmargin\labelsep
	\@nmbrlisttrue\def\@listctr{pubctr}
	\setcounter{pubctr}{#1}\addtocounter{pubctr}{-1}}}
\def\@@publist{\list
	{[\arabic{pubctr}]\hfill}{\settowidth\labelwidth{[999]}
	\leftmargin\labelwidth
	\advance\leftmargin\labelsep
	\@nmbrlisttrue\def\@listctr{pubctr}}}
 \relax
\makeatother
\newskip\humongous \humongous=0pt plus 1000pt minus 1000pt

\newif\ifdtup

\relax

\def\thefootnote{\fnsymbol{footnote}}
\def\a{\alpha}

\def\d{\delta}

\def\f{\phi}

\def\l{\lambda}

\def\q{\theta}

\def\J{\Psi}

\def\S{\Sigma}

\def\Ab{\bar{A}}
\def\gi{g^{-1}}

\def\zb{\bar{z}}

\def\Tb{\bar{T}}
 \def\pp {\partial }
\def\pb {\bar{\partial }}
\def\be{\begin{equation}}
\def\ee{\end{equation}}
\def\ben{\begin{eqnarray}}
\def\een{\end{eqnarray}}

\begin{document}
\renewcommand{\theequation}{\thesection.\arabic{equation}}
\newcommand{\beq}{\begin{equation}}
\newcommand{\eeq}[1]{\label{#1}\end{equation}}
\newcommand{\ber}{\begin{eqnarray}}
\newcommand{\eer}[1]{\label{#1}\end{eqnarray}}
\begin{titlepage}
\begin{center}

\hfill CERN-TH/95-321\\
\hfill ENSLAPP-A-565/95\\
\hfill SNUTP 95-119\\
\hfill hep-th/9512030\\

\vskip .2in

{\large \bf Lagrangian Formulation of\\
Symmetric Space sine-Gordon Models}

\vskip 0.4in

{\bf Ioannis Bakas}
\footnote{Permanent address: Department of Physics, University of
Patras, 26110 Patras, Greece}
\footnote{e-mail address: BAKAS@SURYA11.CERN.CH,
BAKAS@LAPPHP8.IN2P3.FR}\\
\vskip .1in

{\em Theory Division, CERN, 1211 Geneva 23, Switzerland, and\\
Laboratoire de Physique Theorique ENSLAPP, 74941
Annecy-le-Vieux, France\footnote{Present address}}\\

\vskip .3in

{\bf Q-Han Park}\footnote{e-mail address:
QPARK@NMS.KYUNGHEE.AC.KR} and {\bf Hyun-Jong Shin}\footnote{e-mail
address: HJSHIN@NMS.KYUNGHEE.AC.KR}

\vskip .1in

{\em Department of Physics, and
Research Institute of Basic Sciences\\
Kyunghee University, Seoul, 130-701, Korea}

\end{center}

\vskip .4in

\begin{center} {\bf ABSTRACT } \end{center}
\begin{quotation}
\noindent
The symmetric space sine-Gordon models arise by conformal reduction of
ordinary 2-dim $\sigma$-models, and they are integrable exhibiting a
black-hole type metric in target space. We provide a Lagrangian formulation
of these systems by considering a triplet of Lie groups  $F \supset G
\supset H$.
We show that for every symmetric space $F/G$, the generalized sine-Gordon
models can be derived from the $G/H$ WZW action, plus
a potential term that is algebraically specified.
Thus, the symmetric space sine-Gordon models describe
certain integrable perturbations of coset conformal
field theories at the classical level. We also briefly discuss their
vacuum structure, Backlund transformations, and soliton solutions.
\end{quotation}
\vskip0.7cm
December 1995\\
\end{titlepage}
\vfill
\eject
\def\baselinestretch{1.2}
\baselineskip 16 pt
\section{Introduction}
\noindent
Ordinary 2-dim $\sigma$-models were intensively investigated in
the past and their integrability properties are very well
established (see for instance \cite{Adda}, and references therein).
These models are classically conformal invariant,
and therefore the freedom to choose coordinates can be exploited
setting
$T_{zz} = T_{\bar{z} \bar{z}} = 1$ for the components of the classical
stress-energy tensor. This choice was first implemented by
Pohlmeyer to the equations of motion, leading to another class
of integrable systems known as reduced $\sigma$-models
\cite{Pohlmeyer}.
An alternative description was given by
Lund and Regge thinking of the classical equations of motion as
describing the embedding of a 2-dim surface in the target space
of the $\sigma$-model, which in turn is embedded in flat space
\cite{Lund}. Then, the reduction procedure is analogous
to choosing the orthonormal gauge in bosonic string theory, and
the reduced $\sigma$-model describes the dynamics of the
transverse degrees of freedom after solving the constraints.
\footnote{Of course, ordinary $\sigma$-models are not consistent string
backgrounds quantum mechanically; this geometrical approach is
only used to motivate the definition of classical reduced
$\sigma$-models, as it was originally done in the literature.}
The simplest example is the reduction of the
$S^{2}$ non-linear $\sigma$-model, which yields the celebrated
sine-Gordon equation. This method has been generalized to $F/G$
$\sigma$-models following a systematic group theoretical approach
that leads to the so called symmetric space sine-Gordon models
(SSSG) \cite{Eich} \cite{Auria} \cite{Zak}.

The first non-trivial example of such a multi-component generalization
is given by the complex sine-Gordon model, which is obtained from the
$S^3 \simeq SO(4)/SO(3)$ $\sigma$-model by reduction. It involves two
target space fields $\alpha$ and $\beta$ with Lagrangian
\begin{equation}
{\cal L} = \partial \alpha \bar{\partial} \alpha + {\cot}^2 \alpha
\partial \beta \bar{\partial} \beta + {\cos}^2 \alpha .
\end{equation}
Another interesting example of SSSG that arises as reduced
$CP^2 \simeq SU(3)/U(2)$
$\sigma$-model involves three fields $\alpha$, $\beta$ and $\gamma$
with Lagrangian
\begin{equation}
{\cal L} = \partial \alpha \bar{\partial} \alpha + {1 \over 4}
\partial \beta \bar{\partial} \beta + {\cot}^2 \alpha \partial \gamma
\bar{\partial} \gamma + 2 \cos \alpha \cos (\beta - \gamma) .
\end{equation}
All other SSSG models have a well established Lax-pair
formulation, according to the general group theoretical scheme for
reducing the classical equations of motion of 2-dim $\sigma$-models
\cite{Eich} \cite{Auria},
but there has been no Lagrangian formulation known for them up to this day.
The main difficulty is to find a general parametrization of the target
space fields in terms of the Lax-pair variables, so that a Lagrangian
formulation becomes possible. In the known examples this is
described by non-local transformation of variables whose exact form has
not yet been found in the general $F/G$ case.

Our interest in the subject originates from the possible interpretation
of (1.1) and (1.2) as integrable perturbations of conformal field
theory (CFT) coset models. The reduction procedure, which spoils
conformal invariance while preserving integrability, has a rather
dramatic effect in the target space structure of $\sigma$-models.
It is intriguing that the target space metric of the above two
SSSG models is very different from ordinary $\sigma$-models in that there
are singularities reminiscent of CFT black-hole backgrounds in
string theory (although the black-hole interpretation is more
appropriate for non-compact cosets \cite{Ed}). The complex sine-Gordon model
(1.1) has already been considered in some detail
\cite{Bakas} \cite{Park}, providing the
Lagrangian formulation of the parafermionic $SU(2)/U(1)$ coset model
perturbed by its first thermal operator. We will see later that the
reduced $CP^2$ model can also be reformulated as a perturbed
$SU(2) \times U(1) / U(1)$ CFT coset. It will be clear from our description
that the non-local field redefinitions, used in the past to obtain
(1.1) and (1.2) from the corresponding SSSG Lax-pairs, provide the
classical parafermion variables of these CFT cosets.

In this paper we present a systematic Lagrangian formulation of the
SSSG models for symmetric spaces $F/G$, using the gauged
WZW action for $G/H$ plus a suitable deforming potential term.
The general scheme for choosing the triplet of Lie algebras
$({\bf f , g , h})$ with respective associated groups $F \supset G
\supset H$ will be explained in detail. Our construction shows that all
SSSG models
correspond to integrable deformations of certain conformal field theories,
although further work is required to identify the quantum operators
associated with the potential term in all different cases. This approach
certainly serves as a basis for having a completely new look at the
quantum structure of the various multi-component sine-Gordon models.
We apply our approach to the symmetric spaces $F/G \simeq SO(n+1)/SO(n)$,
$SU(n)/SO(n)$, $SU(n+1)/U(n)$, and rederive the Lax-pair formulation
of the SSSG models from the WZW point of view. We also construct a new
class of models based on the symmetric space $Sp(n)/U(n)$. Finally, we
consider the form of Backlund transformations, and present as
example the 1-soliton solution of the $SU(3)/SO(3)$ SSSG model.
Further generalizations of our scheme are also briefly discussed.

We note that for all SSSG models other than (1.1) and (1.2), the
corresponding $G/H$ WZW cosets turn out to have non-abelian group $H$.
It will be clear from our presentation that the required non-local
field redefinitions for having a Lagrangian description of the
underlying Lax-pair equations are actually equivalent to introducing
non-abelian parafermions in the general case, as non-local functions of the
target space fields. It might explain why this
problem was not solved fifteen years ago, without the CFT interpretation
of these models.

\section{The general scheme}
\setcounter{equation}{0}
\noindent
Let  $F/G$ be a symmetric space,  where the Lie algebra decomposition
$\bf{f} = \bf{g} \oplus \bf{k}$ satisfies the commutation relations
\be
[ \bf{g} \ , \ \bf{g} ] \subset \bf{g} \ ,\ [ \bf{g} \ , \ \bf{k} ] \subset
\bf{k} \ , \
[\bf{k} \ , \ \bf{k} ] \subset \bf{g} \ .
\ee
We take two arbitrary elements $T, ~ \Tb $ of $\bf{k}$ and define
${ \bf{h} }$ as the simultaneous centralizer of $T$ and $\Tb $, i.e.
${ \bf{h} } = C_{\bf{g}}(T, \Tb ) = \{ R \in {\bf{g} }: [ R , ~  T]=0 =
[R, ~ \Tb ] \} $.
Then, the Lagrangian formulation of the SSSG model for the symmetric space
$F/G$ is given by the gauged $G/H$ WZW action plus a potential term,
\be
I = I_{WZW}(g) +
{1 \over 2\pi }\int \mbox {Tr} (- A\pb g \gi + \Ab \gi \pp g
 + Ag\Ab \gi - A\Ab ) - I_{P}(g, T, \Tb )
 \ee
where $I_{WZW}(g)$ is the WZW action for a map $g  :
M \rightarrow G \subset F $ of a Lie group $G$ defined on two-dimensional
Minkowski space $M$ \cite{witten}.  The connections $A, ~ \Ab$ gauge the
diagonal subgroup $H$ of $G$. The potential term is given in terms of
$T $ and $\Tb $,
\be
I_{P}(g, T, \Tb ) =  {m^2 \over 2\pi }\int \mbox{Tr}(gT\gi \Tb  ),
\ee
where $m^{2}$ is a mass parameter.

The action (2.2) without the potential term is precisely the action  for
a conformal field theory based on the coset $G/H$ \cite{dimitra}.
Therefore, our model describes an integrable perturbation of
this coset conformal field theory with a deforming potential term
that is  characterized by  $T$ and $ \Tb $ associated with the embedding of
$G$ into $F$. In order to understand the integrability properties of the
model, and make precise connection with the usual formulation of the SSSG
models,
we should write the equations of motion in a zero cuvature form.
Varying the action with respect to  $g$ we have
\be
\d _{g}I = {1 \over 2\pi }\int \mbox{Tr } (- [ ~ \pb + \Ab , ~
\pp + \gi \pp g + \gi A g ~ ] +
m^2 [ ~ T , ~ \gi \Tb g ~ ] )g^{-1}\d g = 0 .
\ee
Since $A$ and $ \Ab $ commute with $T $ and $ \Tb $ we have an identity
$[ ~ \pb + \Ab , ~ T ~ ] = 0 $, and also
\be
\pp ( \gi \Tb g ) + [ ~ \gi \pp g + \gi A g  \ ,  \  \gi \Tb g ~ ] = 0 ,
\ee
which can be combined with (2.4) to give the equivalent zero cuvature
expression
\be
[~ \pp + \gi \pp g + \gi A g  +  \l T \ , \ \pb + \Ab +{m^2 \over \l } \gi
\Tb g ~ ] = 0  ~ ,
\ee
where  $\l $ is an arbitrary spectral parameter. This last equation arises
as the integrability condition of the linear system
\be
(\pp + \gi \pp g + \gi A g  +  \l T )\Psi = 0 \ \ ,\ \
(\pb + \Ab +{m^2 \over \l } \gi \Tb g )\Psi = 0.
\ee
Note that the linear system (2.7) is written utilizing the full algebra
${\bf f}$, whereas the integrability equation (2.6) is only restricted to
the  subalgebra ${\bf g}$ due to the commutation relations (2.1).

The constraint equations arising from the $A $ and $ \Ab $ variations are
\ben
\d _{A}I &=& {1 \over 2\pi }\int \mbox{Tr} ( \ - \pb g
\gi + g\Ab \gi - \Ab \  )\d A
= 0  \\
\d _{\Ab }I &=& {1 \over 2\pi }\int \mbox{Tr} ( \  \gi
\pp g  +\gi A g - A \ )\d\Ab = 0 \ ,
\een
which, when combined with (2.6), yield the zero curvature condition
 $[~ \pp + A , ~ \pb + \Ab ~ ]=0$.
This allows us to fix the gauge, and without loss of generality we set
$A = \Ab =0$ from now on.
Then, (2.6) becomes
\be
\pb ( \gi \pp g ) - m^2 [ ~ T , ~ \gi \Tb g ~ ] = 0,
\ee
and the constraint equations reduce to
\be
( \gi \pp g )_{\bf h } = 0, ~~~~ (  \pb g \gi  )_{\bf h } = 0,
\ee
where the subscript ${\bf h }$ denotes the projection to the ${\bf h}$
subalgebra.

We may solve the equations (2.5) and (2.11) introducing explicit
parametrizations of  $\gi \Tb g$, so that the equation of motion for
$\gi \pp g$, (2.10), becomes a vector type generalization of the usual
sine-Gordon equation. This is actually done  in the next section for
various choices of symmetric spaces. In fact, only these vector type
equations and their zero curvature expressions  are used to define the
SSSG model, as in the earlier works \cite{Eich} \cite{Auria}.
We emphasize, however, that even though the equations of motion can be
expressed solely in terms of the $\gi \Tb g$ variables,
the Lagrangian formulation requires a full parametrization of $g$.
The number of parameters of $g$ is bigger than  $\gi \Tb g$, and the
constraints we have imposed match the difference.
\section{Symmetric spaces}
\setcounter{equation}{0}
\noindent
Compact symmetric spaces have been completely classified by Cartan
(see for instance \cite{helgason}), and they are labeled by type I and
type II. Here, we consider type I symmetric spaces and present explicit
equations for most of them. At the end of the paper, we will make some
comments on the type II spaces, and make connections with other examples of
generalized sine-Gordon models based on $SL(2)$ embeddings \cite{Hollowood}.

\underline{{\bf I. }~  $F/G = SO(n+1)/SO(n) $}
\vglue .1in
We choose $T, ~ \Tb $ and the embedding of $SO(n)$ into $SO(n+1)$ as follows,
\be
T = \Tb = \pmatrix{ 0 & -1 & 0 & \dots&0 \cr
	      1 & 0 & 0 & \dots&0  \cr
	      0 & 0 & 0 &  \cdots  &0 \cr
	      \vdots &&& &\vdots \cr
	       0 && \cdots && 0 }, ~~~
\tilde{g} = \pmatrix{ 1 & 0 & \cdots & 0 \cr
		      0 & && \cr
		      \vdots & & g \in SO(n) & \cr
		      0 &&& } \in SO(n+1 ),
\ee
so that the stability group $H= SO(n-1)$. We also introduce explicit
parametrizations,
\be
\gi \Tb g = \pmatrix{ 0 & V_{0} & \cdots & V_{n-1} \cr
		      -V_{0} & 0 & \cdots & 0 \cr
		       \vdots &\vdots && \vdots \cr
		       -V_{n-1} &0 & \cdots & 0 }
, ~~~
\gi \pp g = \pmatrix{ 0 & 0&0 & \cdots & 0 \cr
				0 & 0 & E_{1} & \cdots & E_{n-1} \cr
				 0  & -E_{1} &&& \cr
				\vdots & \vdots && A = 0 & \cr
				0 & -E_{n-1} &&& } ,
\ee
where we have imposed the constraint (2.11). Then, the elements
$V_{i}=-g_{1, i+1}, $ where $  i = 0,1,...,n-1, $  satisfy the normalization
condition
\be
V_{0}^{2} + \sum_{k=1}^{n-1}V_{k}^{2} = \sum_{k=1}^{n}g_{1k}^{2} = 1.
\ee
The identity (2.5) resolves into component equations,
\be
\pp V_{0} + E_{k}V_{k} = 0 \ , ~~~ \pp V_{i}  - V_{0}E_{i} = 0 ; ~~~
i = 1,2,...,n-1,
\ee
which can be solved for $E_{i}$,
\be
E_{i} = {\pp V_{i} \over \sqrt{1-V_{k}V_{k}}}.
\ee
Then, (2.10) yields the vector type SSSG equation
\be
\pb E_{i} - m^{2}V_{i} \equiv \pb
{\pp V_{i} \over \sqrt{1-V_{k}V_{k}}} - m^{2}V_{i} = 0 ,
\ee
which reproduces the $SO(n+1)/SO(n)$ SSSG model \cite{Eich} \cite{Auria}
from the $SO(n)/SO(n-1)$ gauged WZW theory.
\vglue .2in
\underline{ {\bf II.} ~ $F/G = SU(n)/SO(n)$ }
\vglue .1in
Here, we choose $T, ~ \Tb $
\be
T=\Tb = \pmatrix{ -n+1 &0 & \cdots & 0 \cr
		   0 & 1 && 0 \cr
		   \vdots && \ddots & \cr
		   0 & 0 && 1 },
\ee
and embed the $SO(n)$ group by restricting the $SU(n)$ elements to be real,
so that the stability group $H=SO(n-1)$. We also introduce the
parametrizations,
\be
\gi \Tb g = {\bf 1 }- n \pmatrix{ V_{0} & V_{1} & \cdots & V_{n-1} \cr
		      V_{1} &&& \cr
		       \vdots & & {1 \over V_{0}}V_{i}V_{j} & \cr
		       V_{n-1} & & & } , ~~
\gi \pp g = \pmatrix{ 0 & E_{1} & \cdots & E_{n-1 } \cr
				-E_{1} &  & & \cr
				  \vdots  & &  A= 0 & \cr
				-E_{n-1} &  &  & } ,
\ee
where $i, ~ j$ run from 1 to $n-1$, and  $ V_{i} = g_{1,1}g_{1,i+1}$
satisfy the relation
\be
V_{0}^{2} + \sum_{k=1}^{n-1}V_{k}^{2} = V_{0} .
\ee
The identity (2.5) becomes
\be
\pp V_{0} + 2E_{k}V_{k} = 0 , ~~~
\pp V_{i}  + {1\over V_0}V_{i}V_{k}E_{k} - E_{i} V_0 = 0 ,
\ee
which can be solved for $E_i $,
\be
E_{i} = {1 \over \sqrt{V_0}} \pp (V_{i} \sqrt{V_0} ).
\ee
Changing variables, $V_{0}= 1-U_{k}U_{k} $ and $ V_{i} =
\sqrt{1-U_{k}U_{k}}U_{i} $,
we find that (2.10) becomes the vector type sine-Gordon equation
\be
\pb  { \pp U_{i} \over \sqrt{1-U_{k}U_{k}} }
- m^{2}n^{2}\sqrt{1-U_{k}U_{k}} U_{i} = 0
; ~~~ i = 1,2,...,n-1
\ee
for the $SU(n)/SO(n)$ SSSG model. This corresponds to the maximally
degenerate case of the SSSG models in \cite{Auria}.
The generalization to nondegenerate cases is straightforward choosing
$T = \Tb $ with distinct diagonal elements.
\vglue .2in
\underline{ {\bf III.} $~ F/G = SU(n+1)/U(n) $ }
\vglue .1in
We choose $T, ~ \Tb $ and the embedding of the $U(n)$ group into $SU(n+1)$
as follows,
\be
T = \Tb = \pmatrix{ 0 & -1 & 0 & \dots&0 \cr
	      1 & 0 & 0 & \dots&0  \cr
	      0 & 0 & 0& \cdots  &0 \cr
	      \vdots &&& &\vdots \cr
	       0 && \cdots && 0 }, ~~
\tilde{g} = \pmatrix{ (\mbox{det}g )^{-1} & 0 & \cdots & 0 \cr
		      0 & && \cr
		      \vdots & & g \in U(n) & \cr
		      0 &&& } \in SU(n+1 ),
\ee
so that $H = U(n-1)$. We also have
\be
\gi \Tb g = \pmatrix{ 0 & -V_{0} & \cdots & V_{n-1 } \cr
				V^{*}_{0} & 0  &\cdots  &0 \cr
				  \vdots  &\vdots & & \vdots  \cr
				V^{*}_{n-1} &0  & \cdots & 0 }
, ~~
\gi \pp g  = \pmatrix{ iE_{0} & 0& 0 &\cdots& 0 \cr
		      0 & -iE_{0} & -E_{1}&\cdots & -E_{n-1}  \cr
		       0 &E^{*}_{1} &&& \cr
			\vdots & \vdots  && A=0 & \cr
		       0 & E^{*}_{n-1}& && }
\ee
satisfying the normalization condition
\be
 V_{0}V_{0}^{*} + V_{k}V_{k}^{*} = 1,
\ee
so that  $V_{0} = e^{i\theta }\sqrt{1-V_{k}V_{k}^{*}} $ for some $\theta $.
The identity (2.5) resolves into component equations
\be
\pp V_{i} + iE_{0}V_{i} + E_{i}V_{0} = 0 , ~~~
\pp V_{0} + 2iE_{0}V_{0} - V_{k}E_{k}^{*} = 0,
\ee
which can be solved for $E_{i}$,
\ben
E_{0} &=& {i \over 6V_{k}V_{k}^{*} - 4 }[V_{i}^{*}\pp V_{i} -
\pp V_{i}^{*} V_{i} + 2i(1-V_{k}V_{k}^{*})\pp \theta  ], \nonumber \\
E_{i} &=& -{e^{-i\theta } \over \sqrt{1-V_{k}V_{k}^{*}}}(\pp V_{i} +
iE_{0} V_{i} ) .
\een
Then, the corresponding SSSG equations become
\ben
\pb E_{i} - m^{2}V_{i} &=& 0, \nonumber \\
\pb E_{0} - 2m^{2} \sin{\theta } \sqrt{ 1-V_{k}V_{k}^{*}} &=& 0,
\een
reproducing the previously known result \cite{Eich} \cite{Auria} from
the gauged $U(n)/U(n-1)$ WZW point of view.
\vglue .2in
\underline{{\bf IV.} $ ~ F/G = Sp(n)/U(n) $}
\vglue .1in
In this case, which has not been considered before, we choose
\be
T = \Tb = \pmatrix{ j & 0 & \cdots & 0 \cr
		    0 &  0 & \cdots & 0 \cr
		    \vdots & \vdots && \vdots \cr
		     0 &0 & \cdots & 0 },
\ee
where $j$ is a quarternion satisfying the defining relations $i^2 = j^2 =
k^2 = -1 , \ ij = -ji = k, \ jk = - kj = i , \ ki = - ik = j $. We
embed $U(n)$ into $ Sp(n)$ by restricting the $Sp(n)$ elements to be complex.
Then, the stability group $H = U(n-1)$ and
\be
\gi \Tb g  = j \pmatrix{ V_{0}^{2} & V_{0}V_{1} & \cdots
& V_{0}V_{n-1} \cr
 \vdots &\vdots & & \vdots \cr
V_{n-1}V_{0} &V_{n-1}V_{1} & \cdots & V_{n-1}^{2} }, ~~
\gi \pp g = \pmatrix{ iE_{0} & -E_{1} & \cdots & -E_{n-1 } \cr
				E_{1}^{*} &  & & \cr
				  \vdots  & & A = 0 & \cr
				E_{n-1}^{*} &  &  & },
\ee
where $V_{i} = g_{1,i+1}$ satisfy $V_{0}V_{0}^{*} + V_{k}V_{k}^{*} =1$,
so that $V_{0} = e^{i\theta }\sqrt{1-V_{k}V_{k}^{*}} $ for some $\theta$.
The identity  (2.5) gives
\ben
\pp V_{i} + V_{0}E_{i} &=& 0, \nonumber \\
\pp V_{0} - iE_{0}V_{0} - E_{k}^{*}V_{k} &=& 0,
\een
which can be solved for $E_{0} , ~ E_{i}$,
\ben
E_{0} &=& { -i \over 2- 2V_{k}V_{k}^{*} }[ 2i (1-V_{k}V_{k}^{*} )\pp
\theta  + V_{k}\pp V_{k}^{*} - \pp V_{k} V_{k}^{*}], \nonumber \\
E_{i} &=& -{e^{-i\theta } \over \sqrt{ 1- V_{k}V_{k}^{*}}}\pp V_{i} .
\een
Then, the $U(n)/U(n-1)$ WZW model yields the $Sp(n)/U(n)$ SSSG equations,
\ben
\pb E_{0} + 2m^{2} \sin{2\theta } (1- V_{k}V_{k}^{*} ) &=& 0, \nonumber \\
\pb E_{i}- m^{2} e^{i\theta }\sqrt{1-V_{k}V_{k}^{*}} V_{i} &=& 0,
\een
which clearly differ from case III above, due to the difference in
embedding $U(n)$ in $F$.
\section{Soliton solutions}
\setcounter{equation}{0}
\noindent
Since $T $ and $ \Tb$ commute with $H$, the potential term is also
invariant under the vector gauge symmetry of the gauged WZW action,
\be
g \rightarrow h^{-1}gh \ \ , \ \ A \rightarrow h^{-1}Ah + h^{-1}\pp h \ \ ,
\ \ \Ab \rightarrow h^{-1} \Ab h  + h^{-1}\pb h
\ee
for $h$ valued in $H$. This shows that the potential possesses a flat
direction which is a mere gauge artifact and disappears after the gauge
fixing. In fact, the connection components $A$ and $ \Ab $, due to the lack
of kinetic terms, play the role of Lagrange multipliers,
which impose constraints that suppress the propagating degrees of freedom
along the flat directions. However, there exists a true flat direction
arising from the axial vector symmetry of the potential $I_{P}$,
\be
g \rightarrow hgh \ \ , \ \ A \rightarrow hAh + h \pp h \ \ , \ \
 \Ab \rightarrow h \Ab h  + h\pb h ,
\ee
which in general is not a symmetry of the action (2.2). Thus, this
axial vector symmetry results in a continuous degeneracy of the vacuum.
On the other hand, due to the compactness of the group $G$, the vacuum of
the theory also possesses discrete symmetries, which lead to the soliton
solutions interpolating between two different vacua that are not connected
by the flat directions. Such solutions may be constructed by applying the
dressing method to the linear system of equations
(2.7), or more directly using Backlund transformations.

The Backlund transformation for the symmetric space sine-Gordon models,
in the gauge $A=\bar{A}=0$, is described by
\be
\J_g = {\l \over \l-i \eta} \left( 1+ {m^2 \eta \over \l} g^{-1} M f \right)
\J_f ,
\ee
where $M$ is an arbitrary constant matrix satisfying the condition
\be
[\bar{T}, ~ M] =0 .
\ee
Here $f$, $g$ are elements of $G$, $\l$ is the spectral parameter,
and $\eta \ne 0 $  is a characteristic parameter of the solitons.
$\J_g$ and $\J_f$ satisfy the the linear equations,
\ben
(\pp + \gi \pp g + \l T)\J_{g} = 0 \ , & & \
(\pb + {m^2 \over \l}\gi \Tb g )\J_{g} = 0 ,\nonumber \\
(\pp + f^{-1} \pp f + \l T)\J_{f} = 0 \ , & & \
(\pb + {m^2 \over \l}f^{-1} \Tb f )\J_{f} = 0 .
\een
These equations allow us to eliminate $\J_g $ and $ \J_f $ from (4.3),
leading to the Backlund transformation written only in terms of $g$ and $f$,
\ben
\gi \pp g - f^{-1}\pp f - m^2 \eta [\ \gi M f \ ,
\ T \ ] &=& 0  , \\
\eta \pb (\gi M f) +  \gi \Tb g -
f^{-1} \Tb f &=& 0 \ .
\een
The Backlund transformation offers us the ability to calculate the
1-soliton solution of the theory starting from the vacuum solution, for
which $ f= 1$, and  multi-soliton solutions
using non-abelian superposition rules \cite{Park}.

Consider, for example, the 1-soliton solution
of $SU(3)/SO(3)$.
Specializing to $M^{-1}g = g^{-1} M = 2 P -1$,
so that $P^2=P$, $ P = P^T$, and
setting $f = 1 $, we obtain from (4.6)
\be
2 (2 P-1) \pp P + m^2 \eta \ [\ T\ , ~ 2P-1\ ] =0.
\ee
Multiplying  with $(2 P-1)$ and subtracting the result from
(4.8) we obtain
\be
(1-P)(\pp - m^2 \eta T)P=0 ,
\ee
and similiarly from (4.7) we obtain
\be
(1-P)(\eta \pb - \Tb ) P=0.
\ee
To solve them, it is convenient to represent the projector
$P$ in matrix form as $P_{ij}=
s_i t_j$, where $ i,j=1, 2, 3$, and impose the relation $\sum_i s_i t_i=1$.
The property $P = P^T$ implies $s_i = \a t_i$, where
$\a=s_1^2+s_2^2+s_3^2$. Finally, using equations (4.9) and (4.10), we
may write $s_i$ as follows, using $T = \Tb $ as in (3.7),
\be
s_i = \sum_{j} {(\exp \Sigma T )}_{ij} u_j ; ~~~~
\S =  m^2 \eta  z+  {\zb  \over \eta} ,
\ee
where $(u_1, u_2, u_3 )$ are arbitrary constants
parametrized as $(\cos {\q}, \sin {\q} \cos {\f}, \sin {\q} \sin {\f} )$.
Then, the Backlund transformation yields  the 1-soliton solution of the
equations of motion (3.12), which is summarized as follows,
\ben
E_1 &=& (g^{-1} \pp g)_{12} = 2 m^2 \eta (PT-TP)_{12}
= {3 m^{2} \eta \sin 2\q \cos \f \over \cosh 3\S
-\cos 2\q \sinh 3\S } ,\nonumber \\
V_0 &=& (g^{-1} \Tb g)_{11} = (\Tb -2 \eta \pb P)_{11}
=\left(
-\sinh 3\S +\cos 2\q \cosh 3\S
\over \cosh 3\S -\cos 2\q \sinh
3\S \right)^2 ,\nonumber \\
V_1 &=& (g^{-1} \Tb g)_{12}
={-\sinh 3\S +\cos 2\q
\cosh 3\S
\over {(\cosh 3\S -\cos 2\q \sinh
3\S )^2 }} \sin 2\q \cos \f ,\nonumber \\
E_2 &=&  E_1 \tan \f, \ \ V_2= V_1 \tan \f .
\een

Other examples can be worked out in a similar way, but the computation
becomes technically much more involved.

\vglue .2in
\section{Conclusions and further generalizations}
\setcounter{equation}{0}
\noindent
In this paper we have presented a systematic Lagrangian formulation of
symmetric space sine-Gordon models in terms of the gauged WZW action,
plus a deforming potential term that preserves the integrability of the
system.  Our construction is based on a triplet of Lie groups $(F, G, H)$,
and it has been applied to certain classes of compact symmetric spaces of
type I (in Cartan's classification). In our examples, $T$ and $\Tb$ were
defined using the embedding of $G$ in $F$.
However, the present framework is quite general
and it encompasses other non-abelian generalizations of the sine-Gordon
model as well. For instance, for compact symmetric spaces of type II,
e.g. symmetric spaces of the form
$ G \times G /G$, the elements $g$ and $T$ take the form $g \otimes g$ and
$ T \otimes 1 - 1 \otimes T $ (and similarly for $\Tb$).
The model then becomes effectively equivalent to (2.2),
where $T , ~ \Tb $ belong to the Lie algebra $ {\bf g }$.
One such example is provided by the complex sine-Gordon model, which
arises as reduced $SO(4)/SO(3) \simeq SO(3) \times SO(3) / SO(3)$
$\sigma$-model; it has been described by the $SU(2)/U(1)$ WZW model with
$T = \Tb \in U(1) \subset SU(2)$ \cite{Bakas} \cite{Park}.

Recently, there appeared a new class of generalized sine-Gordon models
based on $SL(2)$ embeddings \cite{Hollowood}.
These systems were also constructed and classified according to a
triplet of Lie groups $(F , G, H)$, where $G$ was chosen
to be the zero graded part of $F$ in
the $SL(2)$ embedding. If we identify $T$ and $\Tb $ with $J_{+} + J_{-}$
of the embedded $SL(2)$ algebra, we find that all of these models (they are
actually five different types) can be incorporated in our SSSG model
construction with symmetric spaces $F/G$ corresponding to the triplets:

\noindent
$AIII \leftrightarrow (SU(2n), ~ SU(n)\times SU(n)
\times U(1) , ~ SU(n))$,

\noindent
$CI \leftrightarrow (Sp(n), ~ SU(n) \times U(1), ~ SO(n))$,

\noindent
$BDI \leftrightarrow (SO(n), ~ SO(n-2)\times U(1) , ~ SO(n-3))$,

\noindent
$DIII \leftrightarrow (SO(4n), ~ SU(2n) \times U(1), ~ Sp(n)) $,

\noindent
$EVII \leftrightarrow (E_{7}, ~ E_{6} \times U(1), ~ H )$,

\noindent
but we have been unable to determine $H$ in the last case.
The interested reader should consult \cite{Hollowood} for further
understanding of the correspondences we are suggesting.

It is clear that the gauged WZW framework we have developed is the most
general for describing various multi-component generalizations of the
sine-Gordon model, using a deforming potential term
$\mbox{Tr}(g T \gi \Tb )$
with appropriately chosen $T$ and $\Tb$ in each case. This suggests a
perturbed conformal field theory approach to the quantization of these
integrable systems, and therefore it is necessary for this purpose to
identify correctly the CFT operators that correspond to the classical
potential terms in the action. It will also be a useful exercise to find
the general form of the non-local field redefinitions that were required
in the old approaches for having a  Lagrangian formulation of SSSG models.
According to earlier work \cite{Bakas}, they should involve the classical
parafermion variables of the corresponding CFT cosets, taking into
account appropriate non-abelian generalizations \cite{Barda}, and express
them non-locally in terms of the target space fields. It will be also
interesting to consider supersymmetric generalizations of SSSG models
\cite{Auria} in our context.
\vskip .2in
\centerline{\bf ACKNOWLEDGEMENT}
\noindent
We thank T. Hollowood and K. Sfetsos for useful discussions in the early
stages of this work. I.B. is also grateful to the CTP of Seoul National
University, the Physics Department of Kyunghee University, and the
Asian-Pacific Center for Theoretical Physics for financial support and
warm hospitality during his visit to Korea. Q.P. and H.J.S. are supported
in part by the program of Basic Science Research,
Ministry of Education BSRI-95-2442, and by Korea Science and Engineering
Foundation through CTP/SNU.

\end{document}